\documentclass[aps,prl,twocolumn,showpacs,preprintnumbers,amsmath,amssymb,superscriptaddress]{revtex4}

\usepackage{graphicx}
\usepackage{dcolumn}

\newcommand{\be}{\begin{equation}}
\newcommand{\ee}{\end{equation}}
\newcommand{\bea}{\begin{eqnarray}}
\newcommand{\eea}{\end{eqnarray}}
\newcommand{\bm}[1]{\mathbf{#1}}
\newcommand{\la}{\langle}
\newcommand{\ra}{\rangle}

\newcommand{\lp}{\left(}
\newcommand{\rp}{\right)}

\def \cF{{\cal F}}

\def \cH{{\cal H}}

\def \tw{{\tilde{\omega}}}
\def \td{{\tilde{\Delta}}}

\newcommand{\ty}[1]{\mbox{\tiny #1}}

\begin{document}

\title{Influence of Disorder on Electron-Hole Pair Condensation in Graphene Bilayers}

\author{R. Bistritzer and A.H. MacDonald}
\affiliation{Department of Physics, The University of Texas at Austin, Austin Texas 78712\\}

\date{\today}

\begin{abstract}
Graphene bilayers can condense into a state with spontaneous interlayer
phase coherence that supports dissipationless counterflow supercurrents.
Here we address the influence of disorder on the graphene bilayer
mean-field and Kosterlitz-Thouless critical
temperatures and report on a simple criteria for the survival of
pair condensation.

\end{abstract}


\maketitle

\noindent
{\em Introduction}--- Superconductivity is a spectacular and
potentially useful condensed matter phenomena. Unfortunately its occurrence
has so far been limited to relatively low temperatures, even in the case of high-$T_c$ materials.
The effective attractive interaction between two electrons that is imperative for superconductivity
relies on some artifice that diminishes the role of the bare strongly repulsive Coulomb interaction.
It has recently been suggested that higher temperature superfluid behavior can be
realized in graphene bilayers when conditions are favorable for Coulomb-driven electron-hole pair formation \cite{us,yogesh}, and that
interesting new superfluid phenomena\cite{jungjung,EisensteinMacDonald} will occur in the
resulting counterflow superfluids. One obstacle which stands in the way of
realizing this state is the inevitable presence of unintended disorder.
In this Letter we derive a simple criterion for the survival of spontaneous interlayer
coherence and conterflow superfluidity in graphene bilayer systems which is informed by
current understanding\cite{tan} of the disorder present in these truly atomic\cite{reviews}
two-dimensional electron systems.

The system we consider\cite{us,yogesh} consists of two parallel graphene layers embedded in a dielectric.
The thin barrier separating the layers suppresses inter-layer tunneling \cite{misorientation}.
External gates can shift the Dirac cones and transfer charge between the layers.
Spontaneous interlayer coherence is most likely to occur\cite{us,yogesh} when
the energetic Dirac-cone shifts are equal and opposite
so there is perfect nesting between the electron Fermi surface of the
high-density layer and the hole Fermi surface of the low-density layer, {\em i.e.} when the bilayer is neutral.
Nearly perfect nesting in neutral bilayers is guaranteed by the nearly-perfect particle-hole
symmetry of graphene's $\pi$-bands.  At sufficiently low temperatures the system is driven to condensation by
the Cooper instability.  Graphene is the ideal system for
bilayer electron-hole condensation\cite{lozovik} because its bands are particle-hole symmetric, because the
lack of a gap assists charge transfer, and because its linear dispersion and truly-2D character
increase electron-hole pairing energy scales at a given carrier density.
The stiffness of the interlayer phase
in the condensed state facilitates counterflow superfluidity and in separately
contacted bilayers supports novel transport phenomena which are still relatively
unexplored\cite{EisensteinMacDonald,jungjung}.

In this work we study the influence of disorder on pair-condensation in graphene bilayers.
We find that disorder suppresses both the mean-field-theory critical temperature
$T_{c}$ and the KT temperature $T_{\ty{KT}}$, but that spontaneous coherence survives when
\begin{equation}
\label{result}
 n_i^2 d^2 \pi \;  \lesssim \; n  \; \lesssim \;  \frac{1}{d^2\pi}.
\end{equation}
Here $n_i$ is the density of the Coulomb scatterers that dominate\cite{nomura,tan} the disorder present
in graphene sheets on dielectric substrates, $n = k_F^2/\pi$ is the external electric field induced carrier density
in each layer, $k_{\ty{F}}$ is the Fermi momentum, and $d$ is the separation between graphene layers.  The right inequality in Eq.(~\ref{result})
expresses the requirement that interlayer interactions be comparable to intralayer interactions discussed
in earlier work\cite{us} while the left inequality places a limit on the allowable disorder strength.
A window in carrier density for spontaneous coherence exists provided that $n_i d^2 \pi \lesssim 1$.  Since
$n_i$ is typically in the range between $10^{11} {\rm cm}^{-2}$ and $10^{12} {\rm cm}^{-2}$ for
graphene on SiO$_2$ substrates,layer separations less than about $10 {\rm nm}$ are
required for spontaneous interlayer phase coherence.
In the body of this Letter we explain
the dependence of $T_c$ and $T_{\ty{KT}}$ on the strength of disorder and
derive condition (\ref{result}).

\noindent
{\em Mean-Field-Theory with Disorder}--- In order to focus on the role of disorder in spontaneous coherence
we simplify the mean-field-theory by neglecting the inessential role of the full valence band of the high-density layer and the
empty conduction band of the low-density layer.
Furthermore we incorporate the effects of the intra-layer Coulomb interaction only through an implicitly renormalized Fermi energy $\epsilon_{\ty{F}}$.
In the same spirit, the momentum dependence of the inter-layer interaction is replaced by an energy cutoff $v/d$ ($\hbar =1$) where $v$ is the velocity of the Dirac quasi-particles.
The influence of disorder is incorporated here using
the self-consistent Born approximation (SCBA).

The basic building block in this
analysis is the retarded Green function matrix $G^r_{\sigma\sigma'}$
which satisfies the Dyson equation
\be
\lp \omega \tau^{(0)} - \epsilon_k \tau^z - \Delta \tau^x - \Sigma \rp G^r = I.     \label{Dyson_eq}
\ee
In Eq.(~\ref{Dyson_eq}) $\epsilon_k = v_F (k-k_F)$ is the band energy, and the Pauli matrices act on
layer labels $\sigma\sigma'$. The disorder and pairing self energies are respectively
\be
\Sigma_{\sigma\sigma'}(\bm{k},\omega) = n_i \sum_{\bm{p}} |S_{\bm{kp}}|^2 Re \left\{ U_{\bm{k}-\bm{p}\sigma}^\star U_{\bm{k}-\bm{p} \sigma'} \right\} G^r_{\sigma\sigma'}(\bm{p},\omega);
\label{Sigma_SCBA}
\ee
and
\be
\Delta = -g \sum_{\bm{k}} |S_{\bm{kp}}|^2 \overline{\la c_{\bm{k}1}^\dagger c_{\bm{k}2} \ra}
\label{Delta_self_consistent}
\ee
where $U_{q\sigma}$ is the disorder potential in layer $\sigma$ and $S_{\bm{k p}} = \lp 1 + e^{i(\theta_{\bm{k}}-\theta_{\bm{p}})} \rp/2$ is graphene's chiral form factor.
The overbar in Eq.(4) denotes the average over disorder and $g$ stands for the inter-layer interaction coupling constant.
Since elastic impurity scattering occurs in the vicinity of the Fermi surface we can assume that
$U=U(\theta)$, {\em i.e.} that the scattering rate depends only on
the angle between the incoming and outgoing momenta.
In a similar spirit we also approximate the density-of-states by its value at the Fermi energy to obtain
\be
G^r(\bm{k},\omega) = \frac{\tw \tau^{(0)} + \epsilon_k \tau^z + \td \tau^x}{\tw^2 - \epsilon_k^2 - \td^2}       \label{Gr}
\ee
where
\be
(\tw,\td) = (\omega,\Delta) + \lp \frac{u}{2\tau_{\ty{S}}},\frac{-1}{2\tau_{\ty{D}}} \rp \frac{1}{\sqrt{1-u^2}},       \label{twd}
\ee
$u = \tw/\td$, and
\be
\frac{1}{2\tau_{\sigma\sigma'}} = \frac{n_i \epsilon_{\ty{F}}}{2 v^2} \int \frac{d\theta}{2\pi} \; Re \left[ U_\sigma^\star(\theta) U_{\sigma'}(\theta) \right]
\; \frac{1+\cos\theta}{2}.   \label{tau_definition}
\ee
We have assumed here that the two layers are identical and set $\tau_{11}=\tau_{22}=\tau_{\ty{S}}$  (S = same layer)
and $\tau_{12}=\tau_{21}=\tau_{\ty{D}}$ (D=different layer).

We now estimate $\tau_{\ty{S}}$ and $\tau_{\ty{D}}$.
The measured transport properties of graphene layers strongly suggest\cite{nomura,tan,hwang} that disorder
is dominated by Coulomb scatterers\cite{nomura,tan} near the interfaces between the dielectric and the graphene sheets.
We therefore expect inter-layer disorder potential correlations to play an inessential role and set
$\tau_{\ty{D}} \to \infty$.  (Including a finite value of $\tau_{\ty{D}}$ would not complicate our analysis in any way.)
After accounting for the difference between transport and scattering lifetimes: $\tau_{tr}/\tau_{S} \approx 2$ \cite{hwang} we follow Ref.[~\onlinecite{tan}] in
estimating that
\begin{equation}
\label{tau}
\epsilon_{\ty{F}} \, \tau_{\ty{S}} \approx 10 \; \frac{n}{n_i}.
\end{equation}
We use this equation below to express $\tau_{\ty{S}}$ in terms of $n_i$.

\noindent
{\em Coherence Criterion}---
There is a complete analogy between the Green function (\ref{Gr}) and that of an electron in a
superconductor with magnetic impurities
once the pair-breaking parameter is identified with $\delta = 1/2\tau_s + 1/2\tau_{\ty{D}}$.
The various ways in which disorder influences coherence in bilayers may therefore
be understood by borrowing results from Abrikosov-Gorkov theory \cite{AG,maki_book}.
In the condensed state the density of states (DOS) of a disordered bilayer graphene is $\nu(\omega)=\frac{\nu_0}{\zeta} \rm{Im} u$ where
$\zeta=\delta/\Delta(T)$ and $\nu_0 = k_F/\pi v$ is the DOS of the normal system.
Disorder smoothes the square root singularity of a clean system. Furthermore it reduces the energy gap to $\Delta(T)[1-\zeta^{2/3}]^{3/2}$ for $\zeta<1$.
For $\zeta>1$ the spectrum is gapless.
The mean field critical temperature $T_c$ in the presence of disorder is given by
\cite{AG}
\be
\log\lp \frac{T_{c0}}{T_c} \rp = \cH(\alpha)     \label{Tc}
\ee
where $\cH(\alpha)=\psi\lp \frac{1}{2} + \frac{\alpha}{\pi} \rp - \psi\lp \frac{1}{2} \rp$ with $\psi$ being the di-Gamma function, $\alpha=\delta/2T_c$,
and $T_{c0}$ is the mean field critical temperature of the clean system.
For weak disorder $T_c = T_{c0} - \pi \delta/4$, {\em i.e.} the critical temperature decreases linearly with $n_i$.
For strong disorder $T_c^2 = 6 \lp \delta/\pi \rp^2 \log\lp \pi T_{c0}/2\delta\Gamma \rp$ where $\Gamma=1.76$.
When $\delta > 0.8 T_{c0}$ the spectrum becomes gapless, however inter-layer phase coherence continues to exist.
Eventually disorder destroys superfluidity altogether when $\delta \gtrsim 0.88 T_{c0}$.  This last result of Abrikosov-Gorkov
theory \cite{AG} can be transformed into a more useful form by noting that for graphene bilayers in the strongly interacting regime ($k_{\ty{B}}=1$)
\begin{equation}
\label{meanfield}
T_{c0} \; \sim \; \frac{0.1 \epsilon_F}{k_Fd}.
\end{equation}
Combining Eq.(~\ref{meanfield})  and Eq.(~\ref{tau}) yields the left inequality in (~\ref{result}).

It is difficult to estimate $T_{c0}$ in the weakly interacting regime when $k_{\ty{F}} d \gg 1$.
In that regime screening and other induced interaction effects are expected to significantly reduce the critical
temperature.  Precise estimates lie beyond the scope of mean field theory and present an interesting
theoretical challenge.  Nevertheless using Eq.(\ref{Tc})
we can find that coherence between the two layers exists as long as
\be
n \gtrsim \lp \frac{v n_i}{10 T_c} \rp^2 e^{-2\cH(\alpha)}.         \label{n_weak_interactions_regime}
\ee
For weak disorder, $\alpha \ll 1$, condition (\ref{n_weak_interactions_regime}) is equivalent to $n \gtrsim  \lp \frac{v n_i}{10 T_c} \rp^2 e^{-\pi \alpha}$.
In the opposite limit ($\alpha \gg 1$) the system will condense for $n \gtrsim  \lp \frac{v n_i}{22 \delta} \rp^2 e^{-\pi^2/12 \alpha^2}$.

\noindent
{\em Kosterlitz-Thouless Temperature}---
In two dimensions superfluidity is destroyed at the KT temperature by vortex anti-vortex proliferation.
To estimate $T_{\ty{KT}}$ we use the Kosterlitz-Thouless equation $\rho_s(T_{KT}) = (2/\pi) \, T_{KT}$
with $\rho_s(T)$ in the presence of disorder calculated from Abrikosov-Gorkov theory.
The phase stiffness is determined from the counterflow current $j_{\ty{D}}$ generated by a uniform gradient of the relative phase between the
two layers and from the relation $2\rho_s A = j_{\ty{D}}(\Delta) - j_{\ty{D}}(\Delta=0)$. The phase gradient $2A$ is obtained
by perturbing the two layers with a pair of constant vector potentials that are equal in magnitude but have opposite signs. The
subtraction of $j_{\ty{D}}(\Delta=0)$ in the last relation is required\cite{us} by a pathology of the Dirac model
which implies that $j_{\ty{D}}$ does not vanish, as required by gauge invariance,  in the normal ($\Delta=0$) state.
We find that
\bea
\rho_s(T) &=& \rho^{(0)}_s - \frac{v^2}{4} \sum_{\bm{k}} \int \frac{d\epsilon}{\pi} n_{\ty{F}}(\epsilon) Im  Tr \left[ G^r(\bm{k} ,\omega) \right]^2  \nonumber \\
&=& \rho^{(0)}_s \big[ 1 + \int d\omega  \; \frac{\partial n_{\ty{F}}}{\partial \omega} \;  \int d\xi  \; \cF(\omega,\xi,\Delta,\zeta)\big] ,
\label{rho_s}
\eea
where $\rho_s(0)=\epsilon_{\ty{F}}/4\pi$ is the zero temperature phase stiffness of a clean system \cite{us} and
\be
\cF(\omega,\xi,\Delta,\zeta) = -\frac{1}{\pi} Im  \int_{-\infty}^\omega d\omega' \frac{\tw^{'2} + \xi^2 + \td^2}{\left[ \tw^{'2} - \xi^2 - \td^2  \right]^2}.
\label{F}
\ee
It is essential that the integration over $\omega'$ in Eq.(\ref{rho_s}) precedes the integration over $\xi$. Vertex corrections to $\rho_s(T)$ vanish
when the disorder potential $U=U(\theta)$ is taken to be independent of the magnitudes of the incoming and outgoing momenta.

In the clean limit $\cF$ is a sum of two delta functions
centered at $\omega = \pm \sqrt{\xi^2+\Delta^2}$. The phase stiffness is then
\be
\rho^{(c)}_s(T) = \rho_s^{(0)} + \frac{v^2}{4}\int d\omega \frac{\partial n_{\ty{F}}}{\partial \omega}  \nu(\omega). \label{rho_s_clean}
\ee
The second term in (\ref{rho_s_clean}) captures the reduction in $\rho_s$ at finite temperature due to the
thermal excitation of quasi-particles. It can be understood as follows.
A uniform phase gradient (induced here by a constant vector potential) generates a
counterflow current in the superfluid. One contribution to the current
$ \sum_{\bm{k}} \bm{v} \ n_{\ty{F}}\lp \epsilon_k + \bm{v A} \rp $ originates from
the change in the occupation of the quasi-particles due to the change of the energy dispersion. To linear
order in $\bm{A}$ the last expression becomes $\frac{v^2}{4} \sum_{\bm{k}} \partial_{\epsilon_k} n_{\ty{F}} \cdot 2A$ from which
the second term in expression (\ref{rho_s_clean})
readily follows. A similar expression holds for the phase stiffness of a BCS superconductor, but with two differences.
First, the velocity of a particle in
a parabolic band is energy dependent and must be left under the integral. Second, the zero temperature phase stiffness naturally arises in a parabolic
band due to the change in the velocity to $\bm{v}+\bm{A}/m$ with $m$ being the effective mass of the particle whereas
in the graphene spontaneous-coherence case it is accumulated at the
Dirac-model's cutoff wavevector, and conveniently captured\cite{us} by
subtracting $j_{\ty{D}}(\Delta=0)$.

The influence of disorder on $\rho_s(T)$ follows mainly from
the finite life time of definite-momentum quasi-particles.
The impurities reduce the energy gap, thus reducing the energy gain due to pair condensation and
correspondingly its sensitivity to phase gradients.
The reduction of the KT temperature due to disorder
is readily obtained from Eq.(\ref{rho_s}) by setting the temperature to $T_{\ty{KT}}$
and using the relation $\rho_s(T_{\ty{KT}})/T_{\ty{KT}} = 2/\pi$.
The pairing potential $\Delta(T_{\ty{KT}})$ must be calculated self consistently
using Eqs.(\ref{Delta_self_consistent},\ref{twd}).

\begin{figure}[h]
\includegraphics[width=0.8\linewidth]{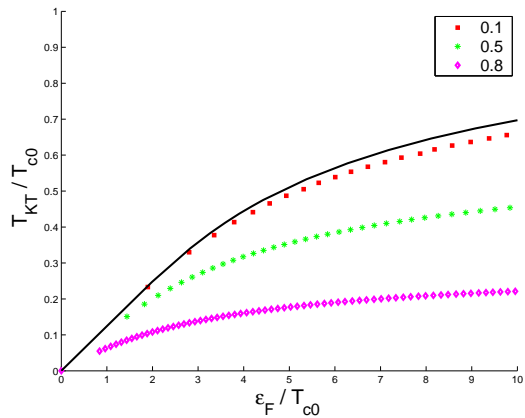}
\caption{(Color online) KT temperature {\em vs.} disorder strength.  Disorder is parameterized
by $\alpha = 1/2\tau_{\ty{S}} T_{c0} = \delta/T_{c0}$.  The solid line in this figure corresponds
to the clean system and the broken lines correspond to (top to bottom) $\alpha=0.1,0.5,0.8$.
In graphene bilayers $\epsilon_{\ty{F}}/T_{c0} \sim 10 k_F d$ in the strongly interacting regime.}
\label{fig:TKT}
\end{figure}

Figure \ref{fig:TKT} shows the reduction of the KT critical temperature of counterflow superfluids due to
disorder.  In this figure
we plot the dependence of $T_{\ty{KT}}$ on the Fermi energy for various strengths of disorder. All energies are scaled
with $T_{c0}$. The strength of disorder is parameterized by $1/2\tau_{\ty{S}} T_{c0} = \delta/T_{c0}$ ($ \sim  n_i k_F d/ n $ in the strongly interacting limit).
In the weakly interacting regime, $\epsilon_{\ty{F}}/T_{c0} \gg 1$ and the KT temperature is of order of $T_{c}$. %(see figure \ref{fig:delta}).
The effect of disorder on $T_{\ty{KT}}$ can then be inferred from its effect on $T_{c}$.
In that regime $\Delta(T_{KT})$ is small hence the disorder will
significantly reduce $T_{\ty{KT}}$. As the interaction increases so does
$T_{c0}$, however the ratio $T_{\ty{KT}}/T_{c0}$ decreases hence $\Delta(T_{\ty{KT}})$ is relatively large.
Thus, in the strongly interacting
regime the effect of disorder on the KT temperature is relatively weak.

\noindent
{\em Discussion}---
Since the pioneering work of Abrikosov, Gorkov and Anderson it has been understood that both the critical temperature and the
order parameter of s-wave superconductors are unaffected by a sufficiently low concentration of non-magnetic impurities \cite{anderson_theorem_localization}. This result, known as
Anderson's theorem \cite{AndersonTheorem}, asserts that even in the presence of (non-magnetic) impurities two states related to one another by
time reversal symmetry will pair and condense. On the
other hand, as experimentally observed \cite{woolf} and theoretically explained by Abrikosov-Gorkov theory,
the values of of $T_c$ and $\Delta$ are suppressed by magnetic impurities
that act to reduce the binding energy of Cooper pairs.
Later work demonstrated that Abrikosov-Gorkov theory applies to other circumstances in which
some perturbation breaks time reversal symmetry, for example thin superconducting
films subjected to a parallel magnetic field \cite{Maki_scB,deGennesTinkham}
or superconductors with an exchange field in the presence of strong spin-orbit interactions \cite{FuldeMaki}.
In this work we extended the range of applicability of Abrikosov-Gorkov theory to dirty graphene bilayers. The extension to semi-conductor based bilayers is trivial.

The analogy between the effects of disorder on superconductivity in metals and on counterflow superfluidity in bilayer systems
is evident if time reversal symmetry in the former is replaced by particle hole symmetry in the latter. Kramer's degeneracy in a non-magnetic superconductor translates to
particle-hole symmetry in the spectrum of the bilayer system. Indeed, if we consider a fictitious scenario in which the disorder
potentials of the two layers are perfectly anti-correlated, {\em i.e.} $U_1(\bm{r}) = -U_2(\bm{r})$, then particle-hole symmetry is preserved,
$\tau_s = -\tau_{\ty{D}}$, the pair breaking parameter $\alpha$ vanishes, and both $T_c$ and $\Delta$ retain their clean system values.
Generic disorder potentials break particle-hole symmetry and therefore provide a particle-hole pair breaking mechanism.
The influence of disorder on the critical temperature is determined by the value of $\alpha=\delta/2T_c$.
Disorder will have a negligible effect on $T_c$ as long as $\alpha \ll 1$
reinforcing the plausibility of condensation in the strongly interacting regime.

Ideas similar to the ones used here may be applied to study disorder effects on exciton condensation in  quantum Hall bilayers.
In current experiments \cite{QHbilayer} $1/2\tau_s T_c$ is of order of unity suggesting that incorporating disorder is imperative for
a correct estimate of the mean field critical temperature and of $T_{\ty{KT}}$.
It is also interesting to consider the influence of magnetic impurities on a bilayer graphene system.
Due to the spin and valley degeneracy of the graphene sheets the exciton pairing is SU(4) symmetric.
Put differently, four identical superfluids coexist in the system. A scattering event by a magnetic impurity
will thus scatter an exciton from one superfluid to another.

\noindent
{\em Acknowledgment ---} We would like to thank the ASPEN center of physics where this work was finalized.
This work has been supported by the Welch Foundation, by the Army Research Office,
by the NRI SWAN Center, and by the National Science Foundation under grant DMR-0606489. RB acknowledges
helpful conversations with H. Fertig and G. Refael.

\end{document}